\documentclass[aps,prd,showpacs,preprintnumbers,amsmath,hyperref,amssymb,floatfix,nofootinbib,12pt]{revtex4-1}

\usepackage{graphicx}
\usepackage{color}
\usepackage{colordvi}
\usepackage[normalem]{ulem}
\usepackage{cancel}
\usepackage{multirow}

\newcommand{\beqn}{\begin{eqnarray}}
\newcommand{\eeqn}{\end{eqnarray}}
\newcommand{\be}{\begin{equation}}
\newcommand{\ee}{\end{equation}}

\newcommand{\ba}{\begin{array}{c}}
\newcommand{\bat}{\begin{array}{cc}}
\newcommand{\ea}{\end{array}}

\newcommand{\bi}{\begin{itemize}}
\newcommand{\ei}{\end{itemize}}

\newcommand{\ts}{\textstyle}
\newcommand{\nn}{\nonumber}

\newcommand{\Int}{\displaystyle{\int}}

\newcommand{\comment}[1]{}
\newcommand{\lp}{\left( }
\newcommand{\rp}{\right) }
\newcommand{\p}{\prime}
\newcommand{\unit}{1\!\!1}

\usepackage[utf8]{inputenc}
\usepackage{xcolor}
\usepackage{slashed}
\usepackage{graphicx}
\usepackage{amsmath}
\usepackage{setspace}

\begin{document}

\title{Contribution of exclusive $(\pi^0\pi^0, \pi^0\eta, \eta\eta)\gamma$ channels to the leading order HVP of the muon $g-2$ }

\author{J. L. Guti\'errez-Santiago}
\email{email: jlgutierrez@fis.cinvestav.mx}
\author{G.~
L\'opez-Castro}
\email{email: glopez@fis.cinvestav.mx}

\affiliation{Departamento de F\'\i sica, Centro de Investigaci\' on y de Estudios Avanzados, Apartado Postal 14-740, 07000 Ciudad de M\'exico, M\' exico }

\begin{abstract}
We evaluate the contributions of $(\pi^0\pi^0, \pi^0\eta, \eta\eta)\gamma$ exclusive channels to the dispersion integral of the leading order HVP of the muon anomalous magnetic moment. These channels are included in some way in previous evaluations of the $\pi^0\omega, \eta\omega$ and $\eta\phi$ contributions to $a_{\mu}^{\rm had, LO}$, where the vector resonances (decaying into $\pi^0/\eta+ \gamma$) are assumed to be on-shell. Since the separation of resonance and background contributions in a given observable is, in general, a model-dependent procedure, here we use pseudoscalar mesons and the photon as the {\it in} and {\it out} states of the  $e^+e^- \to (\pi^0\pi^0, \pi^0\eta, \eta\eta)\gamma$ $S$-matrix, such that the cross section contains the interferences among different contributing to the amplitudes. We find $a^{\rm had, LO}_{\mu}(P^0_1P^0_2\gamma)=(1.13\pm 0.13 ) \times 10^{-10}$, where uncertainties stem mainly from vector meson dominance model parameters.  Improved experimental studies of these exclusive channels in the whole range below 2 GeV would reduce model-dependency.
\end{abstract}

\maketitle

\vspace{-5cm}

\section{ Introduction}  

  During the last two decades, the most accurate measurements of the muon anomalous magnetic moment  $a_{\mu}$ \cite{BNLE821, FNAL2021} have defied an explanation within the standard model (SM) framework.  The reference value of $a_{\mu}$ in the SM prediction \cite{WP}  lies 4.2$\sigma$  below the average value of experimental results $\Delta a_{\mu}=a_{\mu}^{\rm exp}-a_{\mu}^{\rm SM}= 25.1(5.9)\times 10^{-10}$ 
   \cite{PDG}, where theoretical and experimental uncertainties, $4.3$ and $4.1\times 10^{-10}$ respectively, contribute with similar amounts \cite{Davier2017, KNT1, Colangelo2019, Hoferichter2019, DHMZ, KNT2, KURZ2014144, Melnikov2004, Masjuan2017, Colangelo2017, Hoferichter2018, Gerardin2019, BIJNENS2019, Colangelo20200317, COLANGELO201490, Aoyama2012, 
 Marciano2003, Gnendiger2013,  PDG}. The uncertainty in the theoretical value is dominated by input  data used to evaluate the $O(\alpha^2)$ hadronic vacuum polarization (HVP) and also from evaluations of  $O(\alpha^3)$ hadronic light-by-light (H-LbL) contributions. The experimental value includes the recent measurement  of the Muon $g-2$ experiment \cite{FNAL2021}, which is in good agreement with previous results from the BNL 821 collaboration \cite{BNLE821}. Forthcoming experimental results from next runs at Fermilab  as well as J-PARC \cite{Abe:2019thb} and PSI \cite{PSI} will increase the accuracy reducing the current error by up to a factor of three \cite{WP}.

  The recent measurement of $a_{\mu}$ at FNAL \cite{FNAL2021} arrived simultaneously with a new determination of the hadronic contributions based on lattice QCD \cite{Borsanyi2021}. This calculation claims to have reached an accuracy similar to the one of the reference value in the SM  (dispersive calculation of the HVP contributions), but it is closer to the experimental value $\Delta a_{\mu}=a_{\mu}^{\rm exp}-a_{\mu}^{\rm SM, LQCD}= 10.7(6.9)\times 10^{-10}$.  Lattice calculations are performed using the fundamental degrees of freedom of QCD to evaluate the HVP contributions; the dispersive evaluations are built up from the sum of cross sections over exclusive hadronic channels to saturate the HVP in the non-perturbative low energy regime. While dispersive calculations of the HVP contributions using the same input data seems to largely agree among them \cite{WP, Davier2017, KNT1, Colangelo2019, Hoferichter2019, DHMZ, KNT2, KURZ2014144}, new independent and more precise lattice evaluations may confirm or discard the results of Ref. \cite{Borsanyi2021}. 

If more precise evaluations confirm the difference between lattice and dispersive $a_{\mu}^{\rm had}$ results, currently at the 2.1$\sigma$ level, this will become another interesting anomaly to focus attention on theoretical predictions of $a_{\mu}$. One possible explanation to close the gap may be that some missing or poorly measured low-mass hadronic channels in electron-positron collisions contribute to increase the value of the dispersive integral of the HVP. 
 In this paper we study the contributions of the $P_1^0P_2^0\gamma$  processes ($P_{1,2}=\pi$ or  $\eta$ mesons) to the leading HVP contributions of the muon $g-2$ in the SM. These contributions are dominated by a rich structure of resonances with masses below 2 GeV. Actually, some of these resonance contributions like $\omega\pi^0, \phi\eta$ and $\omega\eta$ (with the  subsequent radiative decay of vector mesons) have been included in dispersive evaluations of the leading order HVP \cite{DHMZ, KNT2}, using measurements of the $e^+e^- \to V^0P^0$ cross sections ($V (P)$ will refer hereafter to vector and pseudoscalar mesons). Other exclusive channels involving $\omega/\phi$ resonances as final states have been reported also in dispersive evaluations of the $a_{\mu}^{\rm LO, had}$ \cite{DHMZ, KNT2}.

Strictly speaking, according to the properties of the S-matrix, the amplitudes  involving resonances as incoming/outgoing states are not physical observables \cite{Eden:1966dnq, Stuart:1991xk}: only asymptotic physical states $n$ (not resonances) must be included as intermediate states when saturating the unitarity relation:
\begin{equation}\label{unit}
2{\rm Im}\langle \alpha |T |\alpha \rangle= \sum_n |\langle n| T|\alpha \rangle |^2\ 
\end{equation} 
that stems from the $S$-matrix operator, with $S=1-iT$ and $SS^{\dagger}=\unit$. This unitarity relation  is at the base of the dispersive representation of $a_{\mu}^{\rm had, LO}$ and the hadronic cross sections of $e^+e^-$ annihilations \cite{Bouchiat:1961lbg, Durand:1962zzb, Gourdin:1969dm}.  Therefore, from a theoretical point of view it is not fully consistent to use resonances as physical final states in hadronic $e^+e^-$ cross sections, even though it can be a good approximation, particularly for very narrow resonances (see appendix \ref{Ap1}). This is the main motivation behind the present analysis on $P^0_1P^0_2\gamma$ exclusive channels contributions to $a_{\rm had, LO}$ \footnote{Given their large lifetimes compared to hadrons that undergo dominant strong decays, $\pi^0/\eta$ mesons can be considered asymptotic states.}.

The production cross section of $P_1^0P_2^0\gamma$ states are of the same order in the fine structure constant $\alpha$ as $P^0\gamma$ states, with the later being included in evaluations of  the HVP contribution ($a_{\mu}^{\rm had,LO}(\pi_0\gamma+\eta\gamma)\simeq 5\times 10^{-10}$ \cite{DHMZ, KNT2}).  Note that the corresponding non-radiative $e^+e^-\to P_1^0P_2^0$ channels are not allowed final states, at least at leading order; therefore, $P_1^0P_2^0\gamma$ do not correspond to their photon inclusive processes. One may think  that, given the low threshold for the $\pi^0\pi^0\gamma$ its contributions below the 1 GeV region may be enhanced due to the low energy behaviour of the QED kernel in the dispersion integral for $a_{\mu}^{\rm had, LO}$; however, as it will be shown, the cross sections for $P_1^0P_2^0\gamma$ production is peaked above 1 GeV, leading to suppressed contributions. This property follows from the particular Lorentz structure entering the $\gamma^* \to P_1^0P_2^0\gamma$ vertex which leads to $e^+e^-$ cross sections peaked at center of mass energies above 1.4 GeV. Thus, when those cross sections are inserted into the dispersion integral to evaluate  $a_{\mu}^{\rm had, LO}$, the kernel suppression above 1 GeV can be partially compensated by the effect of heavier resonances. 

 Previous calculations of $e^+e^- \to \pi^0\pi^0\gamma, \pi^0\eta\gamma$ cross sections in the region close to the $\phi(1020)$ meson have been provided in Refs. \cite{Isidori:2006we, Moussallam:2021dpk, Achasov:1987ts, Achasov:2002ir}. The corresponding cross sections measurements were reported in \cite{Achasov:2000ku, KLOE:2002, KLOE:2009ehb}, which focus mainly in the hadron mass distribution in $\phi \to P_1P_2\gamma$ decays. Measurements of the $e^+e^- \to \pi^0\eta\gamma$ cross section in the $\sqrt{s}=1.05-2.0$ GeV region have been reported by the SND collaboration \cite{SND2020}. More recently, the first measuments of the $\eta\eta\gamma$ production cross section were reported in \cite{SND:2021amz}.

In the absence of experimental data (except for the $\pi^0\omega( \to \pi^0\gamma)$ channel \cite{SND2013, SND2016}) in the full range below 2.0 GeV, we base our estimate on a Vector Meson Dominance (VMD) model. This model captures the main features of the dynamics of such processes at energies around the resonance regions, and it can be validated with available data  as is the case with the measured cross section for $e^+e^- \to \pi^0\pi^0 \gamma$ \cite{SND2016}. A more sophisticated treatment of the $\gamma^* \to P_1^0P_2^0\gamma$ vertex can be done in the framework of resonance chiral theory by including the one- ($VP\gamma$) and two-resonances ($VV'P\gamma$) contributions;  Although this analysis is possible it involves a larger set of free parameters associated to the coupling of excited resonances. We do not consider this and other approaches in the  present work.

We organize our paper as follows: after this introduction we describe in Section \ref{Amplitude} the general amplitude and relevant kinematics for the $e^+e^-\to P_1^0P_2^0\gamma$ collisions and introduce some useful notations.  In Section \ref{FFVV} we derive the form factors for the vector-vector  contributions to the hadronic vertex.  Section \ref{a0} considers the vector-scalar contributions in the special case of the $\gamma^* \to \pi^0\eta\gamma$ vertex.   In Section \ref{CS} we use available data on the $e^+e^-\to \pi^0\omega (\to \pi^0\gamma)$ cross section to fit some of the parameters of the model and describe how remaining parameters can be estimated from other data; 
we also provide our results for the cross sections of different channels.  We use the calculated cross sections to compute the dispersive integral and get results for $a_{\mu}^{\rm had, LO}(P_1^0P_2^0\gamma)$ in Section \ref{amu00}. Finally, we give our conclusions in Section \ref{conc} and include two relevant appendices.

\section{\label{Amplitude} Amplitude and kinematics}

In S-matrix theory, the quantum amplitudes describe transitions between incoming and outgoing stable states \cite{Eden:1966dnq}. These initial and final states contain particles which must be described by asymptotic states, {\rm i.e.} free states that can be defined at times long enough before and after the interaction point. According to this tenet of quantum scattering theory, resonances are not asymptotic states; instead, they appear in transition amplitudes as particles described by propagators of unstable states or poles of the transition amplitudes. Physical states also form a complete set $\{ |n\rangle \}$ of states which satisfy the unitarity condition $\sum_n |n\rangle \langle n|=1$. 
The unitarity of the S-matrix operator ($S=1-iT$, where $T$ is the transition operator) implies Eq. (\ref{unit}). 

Similarly, the use of unitarity in the form of the optical theorem, which allows to relate the HVP of $a_{\mu}$ to the cross section for hadron production in electron-positron annihilation via a dispersion relation \cite{Bouchiat:1961lbg, Durand:1962zzb, Gourdin:1969dm}, requires that only asymptotic states are included in the final states of $e^+e^-$ annihilations. Experiments have revealed that multihadron production processes are dominated by intermediate resonances which interfere in the squared amplitude. Owing to interference effects, we can not isolate the observables associated to the production of a given resonance, although it can be a good approximation if the full transition probability is dominated by the production of that resonance \cite{Stuart:1991xk} (see Appendix \ref{Ap1}). One such example is precisely $e^+e^- \to \pi^0\pi^0\gamma$, where the intermediate state $\omega\to \pi^0\gamma$ dominates the cross section. 

In this paper, we study how the cross sections behave when one considers the full $e^+e^-\to P_1^0P_2^0\gamma$ processes including all resonances and their interference and we compare our results with the  particular case where 
 a single resonance contribution is assumed to dominate the cross section. Our purpose is to reevaluate the HVP contribution to $a_{\mu}$ by avoiding the use of resonances as final states.

For definiteness, we introduce the notation $e^+(p_1)e^-(p_2)\to P_1^0(q_1)P_2^0(q_2)\gamma(q_3, \epsilon^*)$, with $p_1^2=p_2^2=m^2,\ q_1^2=m_1^2, \ q_2^2=m_2^2, \ q_3^2=0$ the masses of particles. The square of the center of mass energy is $s=q^2=(p_1+p_2)^2$, such that $s_{\rm min}=(m_1+m_2)^2 \gg 4m^2$.  The final state can be characterized by three Maldestam-like variables $q'^2=(q_2+q_3)^2,\ q''^2=(q_1+q_3)^2$ and $u=(q_1+q_2)^2$, which satisfy the conditions $q'^2+q''^2+u=q^2+m_1^2+m_2^2$ and $q=q_1+q_2+q_3=q'+q''-q_3$ for the energy-momentum conservation. 

\begin{figure}
\begin{center}
\includegraphics[scale=1]{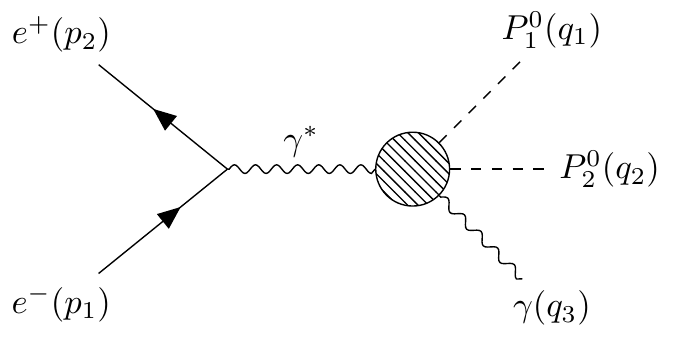} 
\end{center}
\caption{\label{Feynman} Feynman diagram for $e^+e^- \to P_{1}^{0} P_{2}^{0} \gamma$, where $P_{1,2}^{0}=\pi^0$ or $\eta$ .  The bubble represents the effects of strong interactions. 
}
\end{figure}

At the lowest order, the Feynman diagram for this  process is depicted in Figure~\ref{Feynman}. The production amplitude can be presented in the following factorized form:
\beqn \label{AmV}
{\cal M}(e^+e^-\to P_{1}^{0} P_{2}^{0}\gamma)=-ie\frac{\ell^{\mu}}{q^2} H_{\mu}, 
\eeqn
where $\ell^{\mu} = \bar{v}(p_2)\gamma^{\mu}u(p_1)$ is the leptonic current and $H_{\mu}=H_{\mu\sigma}\epsilon^{*\sigma}=\langle P_1^0P_2^0\gamma|j_{\mu}^{\rm em}|0\rangle $ is the hadronic effective current. The most general form of the hadronic tensor 
\begin{equation}\label{genform}
H_{\mu\sigma}= A\left(q\!\cdot\! q_3g_{\mu\sigma}-q_{3\mu}q_{\sigma}\right)+B\left(q'\!\cdot\! q_3g_{\mu\sigma}-q_{3\mu}q'_{\sigma}\right) + C\left(q'\!\cdot\! q_3q_{\sigma}-q\!\cdot\! q_3q'_{\sigma}\right)q'_{\mu}\ ,
\end{equation}
 follows from gauge invariance and Lorentz covariance \footnote{This tensor structure is equivalent to the one given in Eq. (2.5) in Ref. \cite{Isidori:2006we} with different definition of form factors.}. The form factors $A,B,C$ depend upon the independent Lorentz invariants $(q^2, q'^2, q''^2)$ and contains the effects of the strong interactions in the relevant kinematical domain.

The squared amplitude depends upon four independent kinematical variables in addition to $q^2$, which is fixed from the total collision energy. Since the $P_{1}^{0} P_{2}^{0}\gamma$ final states are produced from the $s$-channel one-photon annihilation of $e^+e^-$, the cross section  can be written in the following simple form (see for example \cite{Kubis})
\beqn \label{sig}
\sigma \lp e^+e^- \to P_{1}^0P_{2}^0\gamma \rp = \Int _{\ts m_2^2}^{\ts ( \sqrt{q^2} - m_{1} )^2} \!\!\!\! dq^{\prime 2}\Int_{\ts q''^2_-}^{\ts q''^2_+} dq^{\prime\prime 2}\frac{d^2\sigma}{dq^{\prime 2} dq^{\prime \prime 2}},
\eeqn
where $ \ q^{\p \p 2}_{\pm} =  ( E_{1}^{\star} + E_{3}^{\star})^2 - ( \sqrt{E_{1}^{\star 2} - m_{1}^2} \mp E_{3}^{\star} )^2$,  with $E_{1}^{\star}=(q^2-q^{\p 2}-m_{1}^2)/(2\sqrt{q^{\p 2}})$ and $E_{3}^{\star} = (q^{\p 2}-m_{2}^2)/(2\sqrt{q^{\p 2}})$. The differential cross section in the integrand of Eq. (\ref{sig}) is given by 
\beqn
\frac{d^2\sigma}{dq^{\prime 2} dq^{\prime \prime 2}}=\frac{\alpha }{48 \lp 2\pi \rp^2 q^{6}} \left| H_{\mu\sigma}H^{*\mu\sigma}\frac{}{} \right|.
\eeqn
In the following section we consider the VMD model for the hadronic current. 

\section{\label{FFVV}Form factors of Vector-Vector contributions}

  In the region $\sqrt{q^2} \leq 2$ GeV, the $\gamma^*(q) \to P^0_1(q_1)P^0_2(q_2)\gamma(q_3)$ vertex is dominated by the production and decay of  lowest-lying and excited intermediate resonances.  We will denote with $V (V')$ the intermediate vector resonances as shown in  Figure 2a, ($V, V'=\rho, \omega, \phi$, but we include also their radial excitations for the $s$-channel annihilations). We denote with $S$ the scalar resonances that can mediate the $\pi^0\eta$ final state (see Figure 2b). Accordingly, we can decompose the hadronic tensor into two components
$H_{\mu\sigma}= H_{\mu\sigma}^V+H_{\mu\sigma}^S$, where the superscripts $V$ and $S$ refers to the contributions of diagrams (a) and (b) in Figure 2, respectively.

\begin{figure}
\begin{center}
\includegraphics[scale=1]{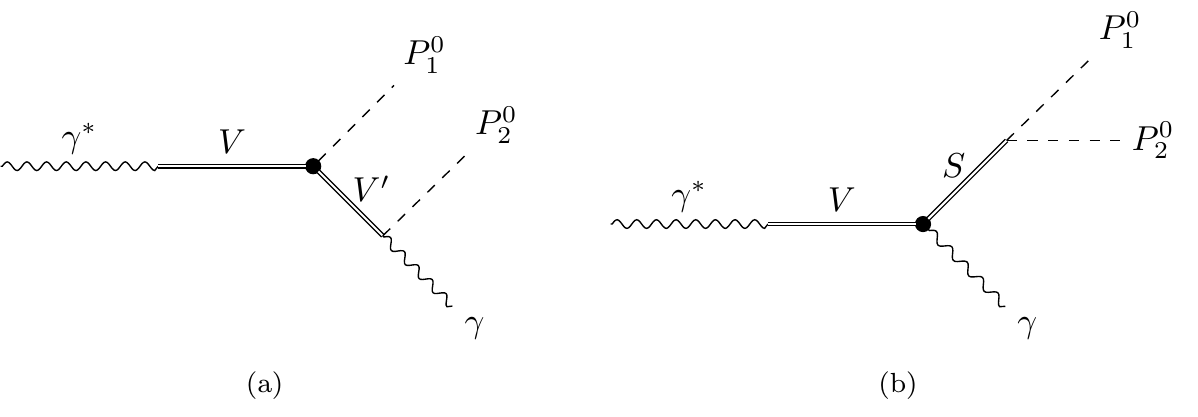} 
\end{center}
\caption{\label{Feynman_a0} Feynman diagrams describing the $\gamma^* \to P_1^0P_2^0\gamma$ vertex in a meson dominance model. Here $V$ (and $V^{\prime}$) are intermediate vector meson resonances and $S$ is a scalar meson. Diagrams with exchanged mesons in the final states for diagram $(\text{a})$ must be added to account for Bose statistics ($P_1^0=P_2^0$) or allowed exchange contributions ($P_1^0\not = P_2^0)$).}
\end{figure}

\subsection{$VV'$ contributions} \label{ffd}

In the VMD model, the contributions to diagrams with two vector resonances are shown in Figure 2a. We need in this case to consider the $V(V')P\gamma$ and $VV'P$ interaction Lagrangians (see Appendix \ref{Ap2}). The phenomenological Lagrangian density required to describe the $VP\gamma$ vertices  is given by \cite{BRAMON2001271}
\begin{equation} 
{\cal L}_{VP\gamma} =g\epsilon_{\mu\nu\alpha\beta}\partial^{\mu}A^{\nu} {\rm Tr}\left[ Q(\partial^{\alpha}V^{\beta}P+P\partial^{\alpha}V^{\beta})\right] \ . \label{lagvpg}
\end{equation}
In this Lagrangian $g$ is a generic coupling, $A^{\mu}$ is the photon field,  $P(V^{\beta})$ is the 3$\times$3 matrix of light pseudoscalar (vector) mesons, and $Q={\rm diag}(2/3, -1/3,-1/3)$ is the matrix of light quark charges. 

Also, one can include isospin and SU(3) breaking effects and try to extract, for lowest lying mesons, the relevant parameters from a global fit to the available data on radiative meson decays, as done for example in Refs. \cite{Escribano:2020} with effective couplings $g_{VP\gamma}$ (which is related to $g$ defined in (\ref{lagvpg}) for each specifc channel \cite{BRAMON2001271, Escribano:2020}). Thus, we can extract the couplings $g_{VP\gamma}$ from the measured rates of radiative meson decays \cite{BRAMON2001271, Escribano:2020}
\be \label{VPg}
\Gamma \lp V\to P\gamma \rp = \frac{1}{3}\Gamma(P\to V\gamma)= \frac{1}{12\pi}g_{VP\gamma}^2\left| \vec{P_{\gamma}} \right| ^3,
\ee
where $g_{VP\gamma}$ is the coupling for the specific $VP\gamma$ decay process and $\vec{P_{\gamma}}$ the photon three-momentum in the decaying particle's rest frame in each specific decay. The values extracted from the radiative decays of light vector mesons are displayed in the lower part of Table~ \ref{TRatios}. 

In addition, we need information on the $VV'P$  couplings of the radially excited vector mesons $V$ 
(here $V'$ is a light vector meson) and its couplings to photons that enter  the $\gamma^*\to V\to V'P$ vertex.  Individual measurements of the strong or lepton-pair decays of excited vector mesons needed to determine those couplings, are not reported by the PDG \cite{PDG}. However, some (model-dependent) analysis of experimental data, mainly from the SND \cite{SND2016, Aulchenko, SND2020, SND2019}, CMD-3 \cite{CMD2020}, BaBar \cite{BaBar2007} and BESIII \cite{BESIII:2020xmw} collaborations, allows to extract the ratio of relevant 
constants $g_{VV'P}/\gamma_V$, where $V$ represents an excited vector meson and $em_V^2/\gamma_V$ its coupling to virtual photons. The relevant product of coupling constants can be extracted from measurements of the cross section at the peak of these resonances which determines the product of their decay rates into $V'P$ and lepton pairs \cite{PDG} through the expression
\be \label{cspeak}
\sigma_{\rm peak}(e^+e^-\to V'P)=\frac{12\pi}{m_V^2}\cdot \frac{\Gamma(V\to e^+e^-)\Gamma(V\to V'P)}{\Gamma_V^2}\ ,
\ee 
where $m_V (\Gamma_V)$ is the mass (width) of the intermediate $s$-channel resonances and $\Gamma(V\to X)$ their partial decay widths into the $X$ channel. The values of the $X_{VV'P}\equiv g_{VV'P}/\gamma_V$ ratios extracted from Eqs. (\ref{cspeak}) and (\ref{ratioStrEl}) are given in the upper part of Table \ref{TRatios}. 

With the above ingredients at hand, we can built the amplitude for $VV'$ contributions of Figure 2a. The hadronic tensor corresponding to the specific configuration shown in that figure 
reads:
\beqn \label{had-ten}
H^V_{\mu\sigma} &=&F^{P_1P_2\gamma} \lp u,q^{\prime 2},q^{\prime\prime 2};q^2 \rp \varepsilon_{\alpha\nu\mu\beta} \varepsilon^{\alpha}_{\ \lambda\sigma\omega} q^{\prime \nu} q^{\beta} q^{\prime \lambda} q_{3}^{\omega}  \nn \\
&=& F^{P_1P_2\gamma} \lp u,q^{\prime 2},q^{\prime\prime 2};q^2 \rp \times \left\{ \frac{}{}q^{\p 2} \left[ \lp q\cdot q_3 \rp g^{\mu\sigma} - q_{3}^{\mu}q^{\sigma}\right]-  \lp q\cdot q^{\p} \rp \left[ \lp q^{\p}\cdot q_3 \rp g^{\mu\sigma} - q_{3}^{\mu}q^{\p \sigma}\right]  \right. \nn \\ 
& & \ \ \left. \frac{}{}
 + \lp q^{\p}\cdot q_3 \rp q^{\p \mu}q^{\sigma} - \lp q\cdot q_3 \rp q^{\p \mu}q^{\p \sigma} \right\} + (q_1\leftrightarrow q_2) \ .
\eeqn
Note that, the last term in Eq. (\ref{had-ten}) symmetrizes the amplitude for identical mesons in the final state ($\pi^0\pi^0, \eta\eta$), and considers the exchange diagram for $\pi^0\eta$ case. In the former case a $1/2!$ factor must be included in the phase space factor. 
In the above expression, the form factor $F^{P_1P_2\gamma} \lp u,q^{\prime 2},q^{\prime\prime 2};q^2 \rp$ contains information on the production and decay of intermediate resonance states. As expected, the hadronic tensor for vector contributions has the structure derived in Eq. (\ref{genform}).

  The squared amplitude for vector contributions will be enhanced at higher c.m.s. energies owing to the quartic momentum dependence Lorentz structure of the hadronic vertex (see Eq. (\ref{had-ten}));  in addition, it will be further enhanced by the effects of radially excited resonances produced in the $s$-channel. Owing to this behavior we will include the light and first/second radially excited $V$ resonances in the $s$-channel, but we keep only the contributions of the lightest vector $V'$ resonances decaying into $(P_{2}^0,P_{1}^0)\gamma$ final states. Accordingly, we write the form factors for the three processes under consideration as follows (the variables $q'^2, q''^2, u$ and $q^2$ in the argument of the form factors are omitted):
\beqn 
F^{\pi^0\pi^0\gamma} &=& F_{\rho}^{\pi^0\pi^0\gamma} + F_{\omega}^{\pi^0\pi^0\gamma} + F_{\phi}^{\pi^0\pi^0\gamma} \ , \label{FF1} \\
F^{\pi^0\eta\gamma}\ &=& F_{\rho}^{\pi^0\eta\gamma} + F_{\omega}^{\pi^0\eta\gamma}  + F_{\phi}^{\pi^0\eta\gamma} \ , \label{FF2} \\ 
F^{\eta\eta\gamma} \ \ &=& F_{\rho}^{\eta\eta\gamma}  + F_{\omega}^{\eta\eta\gamma}  + F_{\phi}^{\eta\eta\gamma}   \ . \label{FF3}
\eeqn
The subindices on the right-hand side refers to the light vector resonances $V'$ decaying into $(P_{2}^0 ,P_{1}^0) \gamma$. An analogous expression to Eq. \eqref{FF2}, namely $F^{\eta\pi^0\gamma}$, must be taken into account for the exchange $\pi^{0} \leftrightarrow \eta$ in the final state. The explicit expressions for each contribution are given in Appendix \ref{Ap2}.

\section{ Form factors for  $e^+e^-\to \pi^0\eta\gamma$ \label{a0}}

In addition to $VV'$ contributions discussed in the previous section, scalar resonances can contribute to $e^+e^-\to P_{1}^{0}P_{2}^{0}\gamma$ as shown in Figure 2b. Among the different final states studied in this paper, only the $a_0(980)$ and (possibly) its `excited' scalar states can contribute sizably as intermediate state in the $\pi^0\eta$ channel  \cite{Achasov:2000ku, KLOE:2002, KLOE:2009ehb, SND2020}.  

The hadronic tensor in this case has a simpler form:
\be
H_{\mu\sigma}^S=\ ie S^{P_1 P_2 \gamma}(q'^2, q''^2, u; q^2) (q\cdot q_3g_{\mu\sigma}-q_{3\mu}q_{\sigma})\ .
\ee
This Lorentz structure is in agreement with the general parametrizacion given in Eq. (\ref{genform}). 

According to Figure \ref{Feynman_a0}b, we need information about the $VS\gamma$ and $S\eta\pi$ interaction couplings. The vertex  $VS\gamma$ responsible for the scalar resonance production is described by the Lagrangian ${\cal L}=(eg_{VS\gamma}/2) F^{\mu\nu}V_{\mu\nu}\phi_{S}$, where $(V, F)_{\mu\nu}$ are the field strenght tensors of the vector-meson and photon, respectively, while $\phi_S$ denotes the field of the scalar meson. The Feynman rule describing the $SP_1P_2$ vertex is given by $ig_{SP_1P_2}$. In terms of these resonances, the form factor for scalar contributions can be parametrized in terms of two scalar resonances $a_0(980), a_0(1450)$ (or $a_0, a'_0$, respectively, for short) decaying into the $\pi^0\eta$ channel, it becomes (this expression agrees with Eq. (4.1) in Ref. \cite{Isidori:2006we} in the case of a single vector and scalar resonance):
\be
S^{P_1 P_2 \gamma}(q'^2, q''^2, u; q^2)=e\sum_V \frac{m_V^2}{\gamma_V D_V(q^2)}\sum_{ S=a_0, a'_0}g_{VS\gamma}\cdot \frac{g_{S\pi\eta}}{D_S(u)} \ .
\ee
In the above expressions, $D_{V,S}(x)=m_{V,S}^2-x-i\sqrt{x}\ \Gamma_{V,S}(x)$ denote the denominators of vector ($V$) and scalar ($S$) resonance propagators, while $\Gamma_i(x)$ denote their total decay widths at squared momentum $x$.


 One may attempt to extract the relevant couplings of scalar mesons from experimental data. Unfortunately, the experimental information on these decays is rather scarce, if not completely missing \footnote{The nature of scalar mesons and their classification is still controversial \cite{PDG}. The resonance parameters and some relevant decay channels of the $a'_0$ are better known than those of the lightest $a_0$ meson. \cite{PDG}.}. Therefore, we will proceed to use a combination of experimental information, theoretical predictions and make the assumption that only one vector resonance in the $s$-channel dominates $a_0, a'_0$ production to provide an estimate of their effects in the cross section:
\begin{enumerate}
\item We will assume that the dominant contribution to the $e^+e^-\to \pi^0\eta\gamma$ cross section below $\sqrt{s}=1.2$ GeV comes from the $\gamma^* \to \phi(1020) \to a_0(980) [\to \pi^0\eta]\gamma$ transition, because both ($\phi$ and $a_0$) can be produced on their mass-shell. Therefore, we use the resonance parameters of the $a_0(980)$ scalar meson as determined in the analysis of the $\phi \to \pi^0\eta\gamma$ hadronic mass distribution measured by KLOE \cite{KLOE:2002} using the resonance model of Ref.  \cite{Isidori:2006we}, namely: 
  $|g_{a_0\eta\pi}|=2.46(14)  \text{GeV}$ and $m_{a_0}=982.5\ \text{MeV}$(fixed), $\Gamma_{a_0}=80\  \text{MeV}$. The coupling $g_{\phi a_0 \gamma}=0.524(11)  \text{GeV}^{-1}$ is extracted from the measured braching fraction of $\phi \to a_0\gamma$ \cite{PDG}.
\item The measured branching fraction of $a'_0(1450)\to \eta\pi$ is reported in \cite{PDG}. Using the $\Gamma(a'_0\to \eta\pi)=(g_{a'_0\eta\pi}^2/8\pi)\cdot |\vec{p}_{\pi}|/m_{a'_0}^2$ decay rate we get $g_{a'_0\eta\pi}=1.46(16)$GeV. The mass and width parameters of the $a'_0$ are taken from the PDG \cite{PDG}.
\item We will assume that, in the region of excited vector $V$ resonances,  only one of them dominates the $\gamma^* \to V \to a'_0\gamma$ vertex. Further, we assume that this vector resonance is the excited state $\phi'=\phi(1680)$. From the following vector-meson dominance relation among the couplings (a similar relation holds for the $a_0\gamma^*\gamma$ coupling)
\be
g_{a'_0\gamma^*\gamma}(q^2) =e\sum_{V }\frac{g_{a'_0V\gamma}}{\gamma_V}\cdot \frac{m_V^2}{D_V(q^2)} \ ,
\ee
and assuming the dominance of the $\phi'(1680)$, one gets at $q^2=0$, $g_{a'_0\gamma\gamma}=eg_{a'_0\phi'\gamma}/\gamma_{\phi'}$. Using the predicted rate for the $\Gamma_{a'_0\gamma\gamma}=(g^{2}_{a'_0\gamma\gamma}/4\pi)m_{a'_0}^3=1.05(5)$ keV \cite{Lu2020}, we get $g_{a'_0\phi' \gamma}/\gamma_{\phi'}=0.0067(2) \text{GeV}^{-1}$ from the above VMD relation \footnote{Note that our $g_{a'_0\gamma\gamma}$ and the one $\widetilde{g}_{a'_0\gamma\gamma}$ used in Ref. \cite{Lu2020}, are related by $g_{a'_0\gamma\gamma}m_{a'_0}^2=\widetilde{g}_{a'_0\gamma\gamma}/2$.}.
\end{enumerate}
  Given all the approximations contained in the derivation of scalar couplings, we must take the predicted effects of scalar mesons in the cross section and $a_{\mu}^{\rm had, LO}(\pi^0\eta\gamma)$ as an indication of their real size.

\section{\label{CS} Cross sections for $P_1^0P_2^0\gamma$ channels}

In this section we consider separately the cross sections for the $e^+e^-\to (\pi^0\pi^0, \pi^0\eta, \eta\eta)\gamma$ reactions. We focus first, in more detail, on the $\pi^0\pi^0\gamma$ channel in order to fix some of the parameters of the model by comparing with available data; this is the channel with the largest cross section among $P_{1}^0P_{2}^0\gamma$ final states owing to the large branching ratio for the $\omega\to \pi^0\gamma$ decay. Thereafter we consider the predictions for the other two channels. 

\subsection{$\pi^0\pi^0\gamma$ final state}
Different experiments have reported measurements of the $e^+e^-\to \pi^0\omega(\to \pi^0\gamma)$ cross section. The SND collaboration has provided results in the energy range $\sqrt{s}=1.05 \sim 2.0$ GeV \cite{SND2013} and $\sqrt{s}=1.047 \sim 2.005$ GeV \cite{SND2016}, while the CMD-2 collaboration in the energy domain 0.920 $\sim $1.380 GeV and DM2 \cite{DM2:1991} in the energy range from 1.350 -- 2.4 GeV. They focus on final states where the $\pi^0\gamma$ system originates from the $\omega(782)$ meson decays which, according to the present discussion, is a part of the full S-matrix amplitude for the $\pi^0\pi^0\gamma$ final state. In the VMD model, the different contributions with intermediate resonances are given by $e^+e^-\to V\to \pi^0 (\rho,\omega,\phi)\to \pi^0\pi^0\gamma$. Accordingly, the general form of the hadronic tensor was given in Eq. (\ref{had-ten}) with the specific invariant form factor 
\be
 F^{\pi^0\pi^0\gamma}(u, q'^2, q''^2; q^2) = F_{\rho}^{\pi^0\pi^0\gamma} + F_{\omega}^{\pi^0\pi^0\gamma} + F_{\phi}^{\pi^0\pi^0\gamma}. \label{ffpipi}
\ee
 The explicit expressions for the different terms are given in Appendix \ref{Ap2}. Dependence upon the same invariant variables must be understood for each term in the right-hand-side of the above equation.

For the denominators  of excited resonances' propagators in Eqs. (\ref{ffpipi}, \ref{ppg1}) we use Breit-Wigner forms with constant widths, namely $D_V(s)=m_V^2-s-im_V\Gamma_V$, where $m_V \ (\Gamma_V)$ denote the mass (width) of resonances. However, following the SND collaboration \cite{SND2013, SND2016} (and our own efforts to achieve a good fit), we use the following expression for the energy-dependent total width  of the $\rho(770)$ meson propagator $D_{\rho}(s)=m_{\rho}^2-s-im_{\rho}\Gamma_{\rho}(s)$:
\be \label{rhow}
\Gamma_{\rho}(s)= \Gamma_{\rho\to \pi\pi}(s)\theta(\sqrt{s}-2m_{\pi})+\Gamma_{\rho \to \omega\pi}(s)\theta( \sqrt{s}-m_{\omega}-m_{\pi})\ ,
\ee
where the energy-dependent partial widths are 
\beqn 
\Gamma_{\rho\to \pi\pi}(s)&=& \Gamma_{\rho} \frac{m_{\rho}^2}{s} \left( \frac{s-4m_{\pi}^2}{m_{\rho}^2-4m_{\pi}^2}\right)^{3/2},  \\
\Gamma_{V\to V'P}&=& \frac{g_{VV'P}^2}{96\pi}\left(\frac{\lambda(s,m_{V'}^2, m_P^2)}{s}\right)^{3/2} \label{wvvp}
\eeqn
with $\theta(x)$ and $\lambda(x,y,z)$  are the step and Kallen functions, respectively. Although Eq. (\ref{rhow}) may look unusual, the opening of new thresholds (like $\omega \pi, \ K\bar{K}, \cdots$) must be included in the decay width as the invariant mass of the resonance increases.

\begin{table}[h!] 
\begin{tabular}{cccc}
\hline
Parameter & Transition &  Value & Reference\\
\hline \hline
& $\ \rho(1450)\to \omega\pi^0$ & $\ 0.5351 \pm 0.0709$ & SND 2016 \cite{SND2016} \\
& $\ \rho(1700)\to \omega\pi^0$ & $\ 0.0425 \pm 0.0207$ & SND 2016 \cite{SND2016} \\
& $\ \omega(1420)\to \rho\pi^0$ & $\ 0.6808 \pm 0.1564$ & SND 2015 \cite{Aulchenko} \\
& $\ \omega(1650)\to \rho\pi^0$ & $\ 0.2329 \pm 0.0286$ & SND 2015 \cite{Aulchenko} \\
& $\ \omega(1420)\to \omega\eta$ & $\ 0.1984 \pm 0.1237$ & SND 2020 \cite{SND2020} \\
& $\ \omega(1650)\to \omega\eta$ & $\ 0.0735 \pm 0.0120$ & SND 2020 \cite{SND2020} \\
$g_{VV^{\p}P}/\gamma_{V}$ & $\ \rho(1450)\to \rho\eta$ & $\ 0.5177 \pm 0.0430$ & \ \ CMD-3 2020 \cite{CMD2020} \\
$\left[\text{GeV}^{-1}\right]$ & $\ \rho(1700)\to \rho\eta$ & $\ 0.0048 \pm 0.0013$ & \ \ CMD-3 2020 \cite{CMD2020} \\
& $\ \phi(1680)\to \phi\eta$ & $\ 0.2875 \pm 0.0818$ & SND 2019 \cite{SND2019} \\
& $\ \phi(2170)\to \phi \eta$ & $\ 0.0048 \pm 0.0074$ & \ \ BaBar 2007 \cite{BaBar2007} \\
& $\ \phi(2170)\to \omega\eta$ & $\ 0.0027 \pm 0.0006$ & \ \ BESIII 2020 \cite{BESIII:2020xmw} \\
\hline 
\multirow{3}{*}{$g_{VVP}/ \gamma_V$ [GeV]$^{-1}$} & $\ \rho \to \rho\eta$ & $\ 1.5181\pm 0.0234$ &  \\
& $\ \phi \to \phi\eta$ & $\ 0.6912 \pm 0.0152$ &  \\
& $\ \omega \to \omega\eta$ & $\ 0.4580 \pm 0.0287$ &  \\
\hline 
& $\ \rho \to \pi^0 \gamma$ & $\ 0.2441\pm 0.0071$ & \cite{Escribano:2020} \\
& $\ \rho \to \eta \gamma$ & $\ 0.4597\pm 0.0174$ &  \cite{Escribano:2020}\\
$g_{VP\gamma}$ & $\ \omega \to \pi^0 \gamma$ & $\ 0.6935\pm 0.0104$ & \cite{Escribano:2020} \\
$\left[\text{GeV}^{-1}\right]$ & $\ \omega \to \eta \gamma$ & $\ 0.1387\pm 0.0087$ &  \cite{Escribano:2020}\\
& $\ \phi \to \pi^0 \gamma$ & $\ 0.0410\pm 0.0037$ & \cite{Escribano:2020} \\
& $\ \phi \to \eta \gamma$ & $\ 0.2093\pm 0.0046$ & \cite{Escribano:2020} \\
\hline
\end{tabular}
\caption{\label{TRatios} Values of model-dependent coupling constants. Entries in the upper part  refers to the values extracted from the peak cross sections of $e^+e^-\to V\to V'P$ as explained in the text using Eqs. (\ref{cspeak}, \ref{ratioStrEl}). Values of the middle  part are extracted using the VMD expressions for $V\to P\gamma$ and $g_{VP\gamma}$ couplings (lower part of Table)  from Ref. \cite{Escribano:2020}.}
\end{table}

The form factors given in Appendix \ref{Ap2} depend upon several parameters: (a) the couplings $g_{VP\gamma}$ needed to describe $(\rho,\omega,\phi)\to P'\gamma$ decays in the sequence $V\to PV'\to PP'\gamma$ are taken from the fits of Ref. \cite{Escribano:2020}, 
and are  listed in the lower part of Table \ref{TRatios}; (b) the ratio of couplings $X_{VV'P} \equiv g_{VV'P}/ \gamma_V$ shown  in the upper part of Table \ref{TRatios} were extracted from experimental values of the peak cross sections (\ref{cspeak})  using the theoretical expression
\be \label{ratioStrEl}
\frac{\Gamma_{V\to e^+e^-}\Gamma_{V\to V'P}}{\Gamma_V^2}=\frac{X_{VV'P}^2\alpha^2}{72}\cdot \frac{\lambda^{3/2}(m_V^2, m_{V'}^2, m_P^2)}{m_V^2\Gamma_V^2}\ ;
\ee
 (c) the strong $VV'\eta$ couplings for light resonances quoted in the middle part of Table \ref{TRatios} were extracted by combining the $g_{V\eta\gamma}$ couplings obtained in \cite{Escribano:2020} and the $em_V/\gamma_V$ couplings for the $\gamma-V$ conversion extracted from measured  \cite{PDG} $\Gamma_{V\to e^+e^-}=(4\pi\alpha^2/3\gamma_{V}^{2})m_V$ partial rates and; (d) the masses and decay widths of remaining radially excited vector resonances were taken from \cite{PDG}. For light vector resonances $\rho/\omega/\phi$ 
 we assume their masses and widhts  world averaged values \cite{PDG}. In the case of the isovector $\rho'=\rho(1450)$ and  $\rho''=\rho(1700)$ mesons we extract the resonance parameters from a fit to the data of the SND collaboration \cite{SND2016}, by assuming their contributions to be complex  relative to the lightest $\rho(770)$ vector resonance.  

In order to be more explicit, we rewrite the dominant contribution as follows:
\be
F_{\omega}^{\pi^0\pi^0\gamma}=ieX_{\rho\omega\pi} \left\{\frac{m_{\rho}^2}{D_{\rho}(q^2)} + \frac{X_{\rho'\omega\pi}e^{i\phi_1}}{X_{\rho\omega\pi}}\frac{m_{\rho'}^2}{D_{\rho'}(q^2)}+ \frac{X_{\rho''\omega\pi}e^{i\phi_2}}{X_{\rho\omega\pi}}\frac{m_{\rho''}^2}{D_{\rho''}(q^2)} \right\} \frac{g_{\omega\pi^0\gamma}}{D_{\omega}(q'^2)}\ ,
\ee
where $X_{V}$  are taken to be real. The $\rho(770)$ meson propagator $D_{\rho}(q^2)$ with the energy-dependent width is given in Eq. (\ref{rhow}). 

\begin{table}[h!] 
\begin{tabular}{ccccc}
\hline
Parameter & SND values \cite{SND2016}& This work & PDG Values \cite{PDG}\\
\hline \hline
$m_{\rho(1450)}$ [MeV] & $1510\pm7$ & $1510\pm 12$ 
& $1465 \pm 25$\\
$\Gamma_{\rho(1450)}$ [MeV] & $440 \pm 40$ & $420\pm 50$
& $400 \pm 60$ \\
$m_{\rho(1700)}$ [MeV]$\dagger$ & $1720\pm20$ & $1720 \pm 20$ 
& $1720 \pm 20$ \\
$\Gamma_{\rho(1700)}$ [MeV] $\dagger$ & $250\pm100$ & $250\pm100$ 
& $250 \pm 100$\\
$g_{\rho\omega\pi}$ [GeV$^{-1}$] & $15.9\pm0.4$ & $17.5\pm 1.3$
& 12.47\\
$\gamma_{\rho}$ & - & - 
& 4.98\\
$X_{\rho'\omega\pi}$ & $0.56\pm 0.05$ & $0.51\pm0.06$
& $ 0.535$ \cite{SND2016} \\
$X_{\rho''\omega\pi}$ & $0.044\pm0.013$& $\quad 0.037\pm 0.012\quad$
& 0.0425 \cite{SND2016} \\
$\phi_{1}$ [deg] & $124\pm17$ & $\quad 114 \pm 34 \quad$
& $127\pm 12$ & - \\
$\phi_{2}$ [deg] & $-63\pm21$ & $-80 \pm 18$
& - \\
$\chi^2/$n.d.f & 0.97 & 0.86 
& -\\
\hline 
\end{tabular}
\caption{\label{Tfit} Results of our fit (third column) to the $e^+e^- \to \pi^0\omega(\to \pi^0\gamma)$ cross section data \cite{SND2016}, compared to results of Ref. \cite{SND2016} and the PDG values \cite{PDG}.  The $\dagger$ symbol means that the parameter has been kept fixed to their PDG values in the fit. }
\end{table}

We can evaluate the cross section using the full S-matrix amplitude by inserting the form factors into Eq. (\ref{had-ten}), taking into account Bose symmetrization terms, and using Eq. (\ref{sig}) for the cross section. In order to compare to available data from SND \cite{SND2016} on the  $e^+e^- \to \pi^0\omega(\to \pi^0\gamma)$ cross section in the $\sqrt{q^2}=1.05\sim 2.0$ GeV region\footnote{We use this dataset because it covers most of the range of center of mass energies. This is the only reason to avoid including data from CMD2 and DM2 collaborations. }, we turn off the first and last terms in Eq. (\ref{ffpipi}).   We let as free parameters: the resonance parameters of the $\rho(1450)$, the complex parameters $X_{\rho'\omega\pi}, X_{\rho''\omega\pi}$ (phases $\phi_1, \phi_2$, respectively) and the $g_{\rho\omega\pi}$ coupling. The third column  in Table \ref{Tfit} collects  results of our fit  to the  cross section data of Ref. \cite{SND2016}. 

\begin{figure} 
\begin{center}
\includegraphics[scale=0.4]{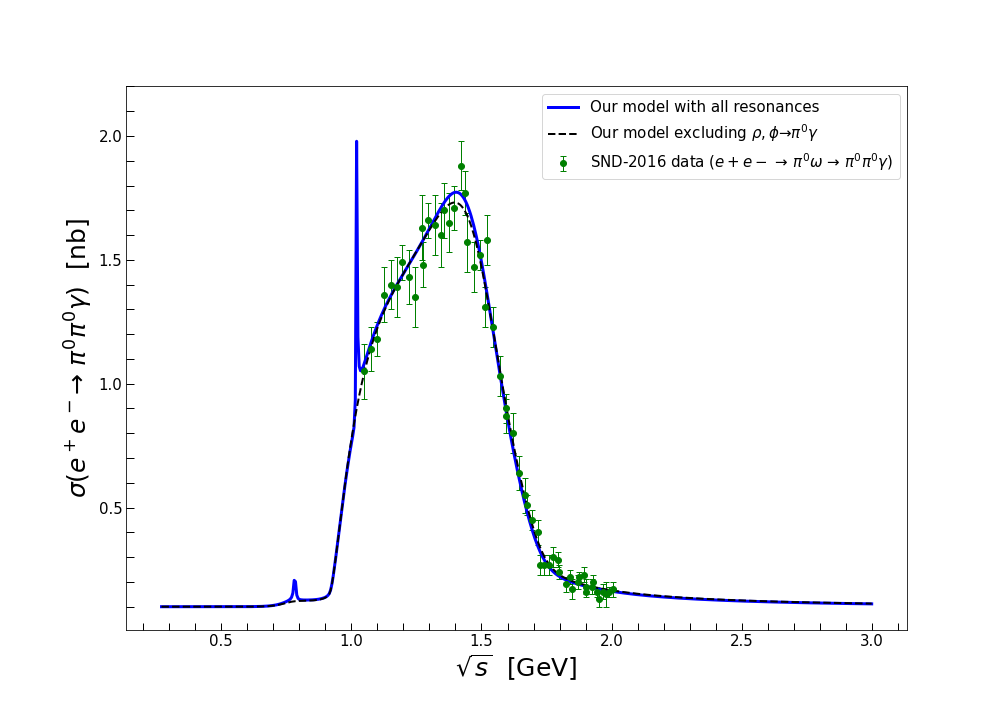} 
\end{center}
\caption{\label{csppg} Cross section for the $e^+e^- \to \pi^0 \pi^0 \gamma$ process. The solid line includes all the resonance contributions given in Eq.  \eqref{ffpipi}. The dashed line corresponds to the dominance of the $\omega\to \pi^0\gamma$ decay,  the second term in \eqref{ffpipi}. The data points correspond to  $e^+e^- \to \pi^0 \omega (\to \pi^0 \gamma)$ measured by SND \cite{SND2016}. }
\end{figure}

A comparison of the second and third columns in the same Table, shows a good agreement between our results and the ones reported by the SND collaboration \cite{SND2016}. Our fit to the experimental data is shown with a dashed line in Figure \ref{csppg}. In the same Figure, we include (solid blue line) the cross section for $\pi^0\pi^0\gamma$ production by taking into account all terms in Eq. (\ref{ffpipi}); except for the narrow peaks at the $\rho(770)$ (suppressed) and at the $\phi(1020)$ (more prominent) resonance positions, the full and $\omega$-dominance contributions agree in all the kinematical range under consideration. 

\subsection{$\pi^0\eta\gamma$ final state}
The amplitude corresponding to $VV'$ contributions (Figure \ref{Feynman_a0}a) for $e^+e^-\to \pi^0\eta\gamma$ must be added with the diagram arising from the exchange of $\pi^0$ and $\eta$ mesons in the final state. Note however that since the intermediate resonances $V$  and $V'$ must be chosen to conserve strong interaction symmetries in the  $V\to \eta V'$ and $V\to \pi^0 V'$ vertices. Since this exchange contribution does not correspond to the exchange of identical particles in the final states, we do not have to add a 1/2! factor in the phase space. 

As it was discussed in Section \ref{a0},  contributions mediated by scalar mesons can appear in the  $\pi^0\eta$ system through the $e^+e^- \to \gamma S(\to \eta\pi)$ mechanism $(S=a_0,a'_0)$. Unfortunately, the situation concerning the experimental information on decay properties of the $a_0(=a_0(980))$ resonance and its nature as a $q\bar{q}$, tetraquark or molecular state  is not very clear so far \cite{PDG, Scalar1, AMSLER200461, BUGG2004257, KLEMPT20071}.  In contrast, the corresponding information for the $a'_0(=a_0(1450))$ properties is better known \cite{PDG}. 

Despite these limitations, we have attempted an estimate of the effects of scalar resonances. We assume that the dominat contribution is given by the $\gamma^* \to \phi(1020)\to a_0\gamma$ chain contribution.  Similarly, we assume that the $a'_0=a_0(1450)$ production is dominated by $\gamma^*\to \phi'(1680)\to a'_0\gamma$. Our assumptions  are based on the fact that at these center of mass energies both, the $(\phi,\phi')$ vector  and the $(a_0,a'_0)$ scalar resonances can be produced on-shell, giving the largest contributions to the cross sections. Values of the coupling constants required in the model were described in Section \ref{a0}.  Of course it corresponds to experiments to resolve the resonant structures present in the $\eta\pi^0$ and $s$-channels in the energy region under consideration.

\begin{figure}
\begin{center}
\includegraphics[scale=0.4]{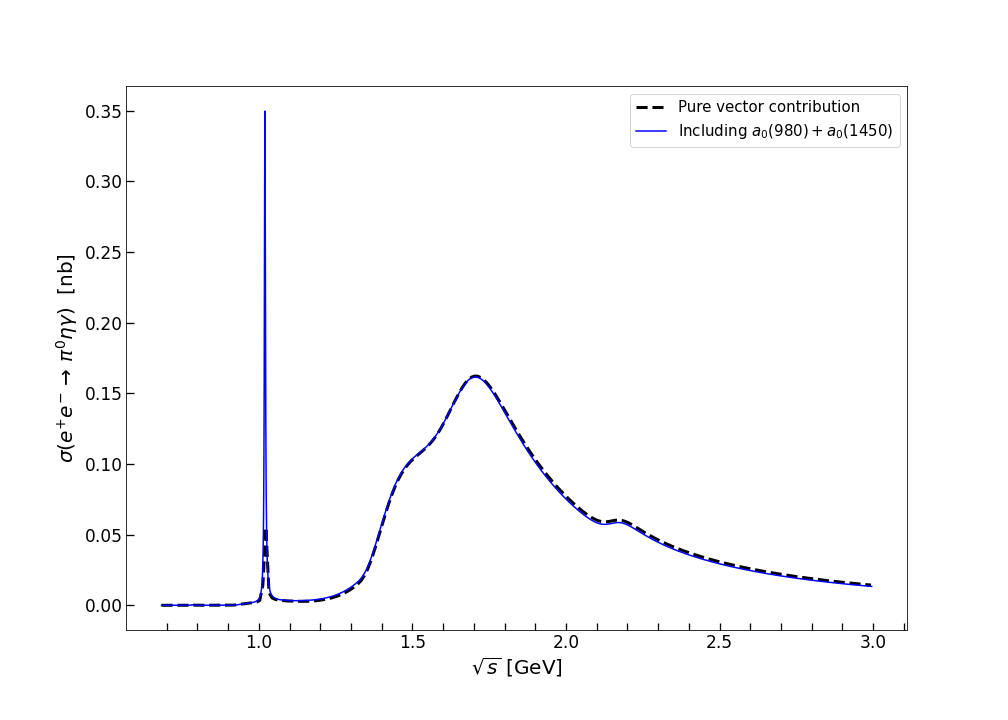} 
\end{center}
\caption{\label{cspega0} Cross section for the $e^+e^- \to \pi^0\eta\gamma$ process. The dashed line corresponds to the pure $\left(V,V^{\prime}\right) = \left(1^{--},1^{--}\right)$ contributions. The solid line includes, in addition, the effects of the $a_0(980)$ and $a'_0(1450)$ scalar mesons.
}
\end{figure}

The  plots of the cross section are given  in Figure \ref{cspega0} as a function of the center of mass energy. The 
continuous line represents the sum of all the contributions, while the dashed 
line corresponds to the pure vector-vector ($V,V'$) contributions. The sharp peak observed to the left is the effect of the $\phi$ meson decaying into the $a_0(980)$ meson and a photon; 
since the $\phi$ is a very narrow resonance, its contribution to $a_{\mu}^{\rm had, LO}(\pi^0\eta\gamma)$ is subdominant. 
On the other hand, the effects of the $a_0(1450)$ scalar meson will be suppressed in $a_{\mu}^{\rm had, LO}$ given the falling of the QED kernel in the dispersive relation.

\subsection{$\eta\eta\gamma$ final state}

The threshold energy for the $e^+e^-\to \eta\eta\gamma$ is  $\sqrt{s}\approx 1.096$ GeV, well above the region of light vector resonances in the $s$-channel. The form factors for this final state are given in 
Eq. (\ref{FF3}) and (\ref{ppg1}), and the hadronic tensor (\ref{had-ten}) must include a symmetrization term according to Bose statistics. Using the input couplings shown in Table \ref{TRatios}, and the convention for the propagators discussed in previous sections, we evaluate the cross section using Eq.~(\ref{sig}). 

\begin{figure}
\begin{center}
\includegraphics[scale=0.4]{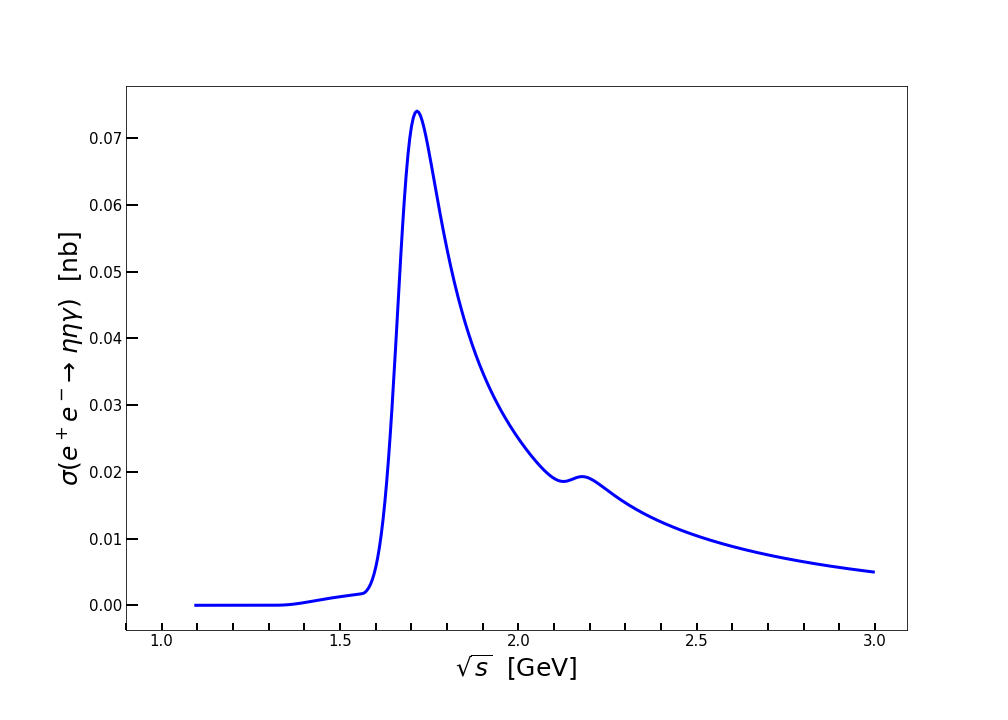} 
\end{center}
\caption{\label{cseeg} Cross section for the $e^+e^- \to \eta \eta \gamma$ process. We use the VMD parameters reported in Table~\ref{TRatios}.
}
\end{figure}

  In Figure \ref{cseeg} we plot the $e^+e^-\to \eta\eta\gamma$ cross section from threshold up to 3.0 GeV. In the absence of experimental information on this decay channel, we assume the different contributions to add coherently with real and positive couplings between different resonance contributions\footnote{Given that $\eta\eta\gamma$ contribution to $a_{\mu}^{\rm had, LO}$ is at the level of $10^{-12}$, for now we can keep this approximation as safe.}. A dominant peak due to the $\rho(1700)$ is observed and a smaller peak is barely visible at the $\phi(2170)$ resonance position.  
As expected, the cross section for $\eta\eta\gamma$  smaller than the one due to $\pi^0\pi^0\gamma$ and $\pi^0 \eta\gamma$.

\section{\label{amu00}$P^0_1P^0_2\gamma$ contributions to $a_{\mu}$}

The HVP contributions to $a_{\mu}$ due to $e^+e^-\to P^0_1P^0_2\gamma$ processes can be written as follows \cite{PhysRev.168.1620, PhysRev.174.1835}
\be \label{amu-disp}
a_{\mu}^{\rm HVP, LO}(P^0_1P^0_2\gamma) = \left(\frac{\alpha}{3\pi}\right)^2 \int_{(m_1+m_2)^2}^{\infty} ds \frac{K(s)}{s}\frac{ \sigma(e^+e^- \to P_{1}^0P_{2}^0\gamma)}{\sigma_{\rm pt}}\ ,
\ee
where $\sigma_{\rm pt}$ is the point cross section for muon-pair production and $K(s)$ is  the QED Kernel that can be found, for instance, in Ref. \cite{WP}.

If we insert the cross sections evaluated in this work into Eq. (\ref{amu-disp}), we get the values shown in the second column of Table \ref{Results}. The second of the two results indicated  for the $\pi^0\eta\gamma$ contribution corresponds to the inclusion of scalar resonances in this channel. Our largest uncertainty appears in the $\pi^0\pi^0\gamma$ contribution. This stems mainly from the uncertainties in the fitted $\rho'\omega\pi$ coupling quoted in Table \ref{Tfit}  \cite{PDG}. Since we do not use the dataset of all $e^+e^- \to \pi^0\omega(\to \pi^0\gamma)$ experimental cross sections, our quoted uncertainty for the $a_{\mu}^{\rm had, LO}(\pi^0\pi^0\gamma)$ channel basically  turns out to be larger that the ones quoted by references \cite{DHMZ, KNT2} (see discussion below).

We can attempt to make a (risky) comparison with Refs. \cite{DHMZ, KNT2}, who have provided the evaluations of the $\pi^0\omega(\omega\to \pi^0\gamma)$, $\eta\omega$ and $\eta\phi$ contributions. For the values of $a_{\mu}^{\rm had, LO}$ for the latter two channels provided in Refs. \cite{DHMZ, KNT2}, we add the subsequent decays of $(\omega, \phi)$ mesons into $(\pi^0, \eta)\gamma$ decays, which is justified in Appendix A. Under these assumptions we can estimate the $P_1^0P_2^0\gamma$ contributions as follows ($B(X)$ denotes the branching fraction for channel $X$):
\beqn \label{ABR}
a_{\mu}(\pi^0 \eta \gamma) &\simeq & a_{\mu}(\eta \omega)\cdot B(\omega \to \pi^0 \gamma) + a_{\mu}(\eta \phi)\cdot B(\phi \to \pi^0 \gamma), \nonumber \\ 
a_{\mu}(\eta \eta \gamma) &\simeq & a_{\mu}(\eta \omega)\cdot B(\omega \to \eta \gamma) + a_{\mu}(\eta \phi)\cdot B(\phi \to \eta \gamma). 
\eeqn
Clearly, this represents, at most, an approximation to the complete evaluation. We use the values $a_{\mu}^{\rm had, LO}(\eta\omega)= 0.35(1)(1)\ [0.30(2)]$ and $a_{\mu}^{\rm had, LO}(\eta\phi)= 0.33(1)(1)\ [0.41(2)]$ from Ref. \cite{DHMZ} (values obtained in \cite{KNT2} are indicated within square brackets), all numbers in $10^{-10}$ units, and the branching ratios reported in \cite{PDG} for the radiative decays of vector mesons.

\begin{table}[h!]
\begin{tabular}{ccccc}
\hline
& & \multicolumn{3}{c}{$a_{\mu}^{\rm had, LO} (X) \times 10^{-10}$} \\ \cline{3-5}
X Channel & Contributions & This work & DHMZ \cite{DHMZ} & KNT \cite{KNT2}\\
\hline \hline
&&&&\\
$\pi^0\pi^0\gamma$ &  $(V, V')$ & $1.002^{+0.129}_{-0.136}$ 
& $\quad 0.94(1)(3)\quad $ & $0.88(2)$\\
 &&&&\\
$\pi^0\eta\gamma$ &  $(V, V')$ & $0.086^{+0.002}_{-0.001}$
& $\quad 0.030(2) \quad$ & $0.026(2) $ \\ \\
$\pi^0\eta\gamma$ & $\quad (V, V')+a_0+a'_0 \quad$ & $0.087\pm 0.001$
 && \\
 &&&&\\
 $\eta\eta\gamma$ & $(V,V')$ & $0.0431^{+0.0001}_{-0.0002}$
& $\quad 0.0045(2) \quad$ & $0.0055(3)$\\
 &&&&\\
\hline
 &&&&\\
& $ (V,V')$ & $1.131^{+0.129}_{-0.136}$
&&\\
 &&&&\\
  & $ (V,V')+a_0+a'_0$ & $1.132^{+0.129}_{-0.136}$
  &&\\ 
  &&&&\\
\hline 
\end{tabular}
\caption{\label{Results} Contributions of $X=P_1P_0\gamma$ exclusive channels to $a_{\mu}^{\rm had, LO}$ (in $10^{-10}$ units).  The results of this work are given in the third column. 
 Columns fourth and fifth for the $(\pi^0\eta,\eta\eta)\gamma$ contributions refer to the values `estimated' according to Eqs. (\ref{ABR}) from the values of $\eta\omega$ and $\eta\phi$ contributions reported  in Refs. \cite{DHMZ} and \cite{KNT2}, respectively. $(a_0, a'_0)$ denote the isovector scalar mesons.}
\end{table}

In columns fourth and fifth of Table \ref{Results} we write the values `estimated' following the above  procedure. These values are underestimated with respect to our results and, in the case of the $\eta\eta\gamma$ channel, by almost one order of magnitude. This is expected since  $\rho\eta$ and $\rho\pi^0$ exclusive channels are not reported separately and interferences are neglected in Eqs.~(\ref{ABR}).


\section{\label{conc}Conclusions}

In this paper we have considered the contributions of the neutral $e^+e^- \to \pi^0\pi^0\gamma, \ \pi^0\eta\gamma$ and $\eta\eta\gamma$ exclusive channels to the leading order HVP contributions of the muon anomalous magnetic moment. We evaluate these contributions by considering the full S-matrix amplitude for transitions between these asymptotic states, without cutting intermediate resonances. These decays are not the photon-inclusive channels of $e^+e^- \to P_1^0P_2^0$, $P_{1,2}=\eta$ or $\pi$, because such transitions are not allowed (at least at the lowest order in $\alpha$) and are expected to be of the same order in $\alpha$ as the $\pi^0\gamma$ and $\eta\gamma$ channels. As it is well known \cite{DHMZ, KNT2}, the later contribute close to 1\%  to the total contributions of $a_{\mu}^{\rm had, LO}$.

We describe the $\gamma^* \to P_1^0P_2^0\gamma$ vertex in the framework of Vector Meson Dominance model. We validate this particular model by fitting the available data on the $e^+e^-\to \pi^0\omega(\omega\to \pi^0\gamma)$ channel.  From the calculated cross sections we evaluate the corresponding dispersion integral and get the following prediction: 
\be
a_{\mu}^{\rm had,LO}(\pi^0\pi^0\gamma+\pi^0\eta\gamma +\eta\eta\gamma) = (1.13^{+0.13}_{-0.14}) \times 10^{-10}.\nn
\ee
This result is dominated by the $\pi^0\pi^0\gamma$ exclusive channel; this is in reasonable good agreement with the evaluation of Refs. \cite{DHMZ, KNT2} for the $\pi^0\omega(\omega \to \pi^0\gamma)$, where a comparison is possible. The other two contributions are more suppressed and a comparison with existing calculations is not straightforward. Our quoted uncertainty is dominated by errors in the strenght coupling of the $\rho'\to  \pi^0\pi^0\gamma$ decay within the VMD model and the particular dataset of $e^+e^- \to \pi^0\omega(\to \pi^0\gamma)$ measurements \cite{SND2016} used in our analysis.

The cross sections for $P_1^0P_2^0\gamma$ production are peaked in the region populated by excited vector resonances in the $s$-channel.  This introduces important uncertainties in the calculation as long as the information on the parameters and decay properties of excited resonances are rather scarce or not very well known.  In order to avoid all the uncertainties related to a particular model, it would be necessary to have better experimental data for these $P_1^0P_2^0\gamma$ final states in electron-positron collisions in the region below 2 GeV. 

  Of course, this dispersive calculation of $a_{\mu}^{\rm had, LO}(P_{1}^0P_{2}^0\gamma)$ presented in this paper does not contribute sizably to close the gap with the measured \cite{BNLE821, FNAL2021} and the lattice calculations of Ref. \cite{Borsanyi2021}. We address the problem of using exclusive channels with resonances and using them as inputs in the evaluation of $a_{\mu}^{\rm had, LO}$.  Using the  S-matrix formalism with asymptotic states is important to asses the size of approximations done when one consider resonances as on-shell states and neglects interference with other contributions to the amplitude. It may not be obvious that separating resonance and background contributions from measured observables, is just an approximation. The most clear example that shows that  interference effects are important is frequently found in the PDG \cite{PDG}, where the sum over final states involving resonances sometimes exceeds the branching ratios for some specific channels (for example $B(D^0\to \pi^+\pi^-\pi^0)=(1.49\pm 0.06)\%$ while  $\sum_{i,j}B(D^0\to (\rho^i(770)\pi^j)^0\to \pi^+\pi^-\pi^0)=(1.91\pm 0.05)\%$ \cite{PDG}).

\appendix
\section{\label{Ap1} Three Body Scattering Processes}

Consider the scattering process $A+B\to C+D+E$. In order to illustrate our point, consider that there are two contributions to the amplitude: ${\cal M}={\cal M}_R+{\cal M}_B$, where the subindex $R$ refers to the production and decays of a resonance $R$: $A+B\to C+R(\to D+E)$ and the subindex $B$ refers to a background (which also may include another less prominent resonance). Accordingly, the cross section contains three terms:
$\sigma =\sigma_R+\sigma_B+\sigma_{\rm int}$  ,
where the subdindex  `int' refers to the interference of resonant and background amplitudes. 

Isolation of the resonant cross section from the full observable is not possible in general since this requires a good control of background terms. Furthermore, if gauge invariance and gauge-independence is not satisfied by individual contributions in the S-matrix amplitude, the resonance cross section can keep residual gauge-dependence \cite{Stuart:1991xk}. Note that if background contributions are negigible small in the region around  $(p_C+p_D)^2\approx m_R^2)$, one can approximate $\sigma(A+B\to C+D+E) \approx \sigma(A+B\to A+R)\cdot {\rm BR}(R\to D+E)$, where the last factor denotes the branching fraction for the $R\to D+E$ decay. This seems to be the case of the $\pi^0\pi^0\gamma$ channel discussed in the present paper, where the calculations of the cross section using the full S-matrix for asymptotic states and the aproximation corresponding to $\sigma(e^+e^- \to \omega(\to \pi^0\gamma)\pi^0)$ give very similar results.

\section{\label{Ap2} Form Factors in $P_{1}^0P_{2}^{0 }\gamma$ Transitions} 

In this appendix we provide the expressions for the  form factors that contribute to the $P_{1}^0P_{2}^0\gamma $ transitions as defined in Section \ref{ffd}

\beqn  \label{ppg1}
F^{\pi^0\pi^0\gamma}_{\rho}&=&ie\sum_{V=\omega,\phi, \cdots} \frac{g_{V\rho^0\pi^0}}{\gamma_V}\cdot \frac{m_V^2}{D_V(q^2)} \cdot \frac{g_{\rho^0\pi^0\gamma}}{D_{\rho}(q'^2)}  \nonumber \\
F^{\pi^0\pi^0\gamma}_{\omega}&=&ie\sum_{V=\rho,\rho', \cdots} \frac{g_{V\omega\pi^0}}{\gamma_V}\cdot \frac{m_V^2}{D_V(q^2)} \cdot \frac{g_{\omega\pi^0\gamma}}{D_{\omega}(q'^2)} \nonumber \\
F^{\pi^0\pi^0\gamma}_{\phi}&=&ie\sum_{V=\rho,\rho', \cdots} \frac{g_{V\phi\pi^0}}{\gamma_V}\cdot \frac{m_V^2}{D_V(q^2)} \cdot \frac{g_{\phi\pi^0\gamma}}{D_{\phi}(q'^2)}  \nonumber \\
F^{\pi^0\eta\gamma}_{\rho}&=&ie\sum_{V=\omega,\phi, \cdots} \frac{g_{V\rho^0\pi^0}}{\gamma_V}\cdot \frac{m_V^2}{D_V(q^2)} \cdot \frac{g_{\rho^0 \eta \gamma}}{D_{\rho}(q'^2)}  \nonumber \\
F^{\pi^0\eta\gamma}_{\omega}&=&ie\sum_{V=\rho,\rho', \cdots} \frac{g_{V\omega\pi^0}}{\gamma_V}\cdot \frac{m_V^2}{D_V(q^2)} \cdot \frac{g_{\omega \eta \gamma}}{D_{\omega}(q'^2)} \nonumber \\
F^{\pi^0\eta\gamma}_{\phi}&=&ie\sum_{V=\rho,\rho', \cdots} \frac{g_{V\phi\pi^0}}{\gamma_V}\cdot \frac{m_V^2}{D_V(q^2)} \cdot \frac{g_{\phi \eta \gamma}}{D_{\phi}(q'^2)} \nonumber \\
F^{\eta \pi^0\gamma}_{\rho}&=&ie\sum_{V=\rho, \rho' \cdots} \frac{g_{V\rho^0\eta}}{\gamma_V}\cdot \frac{m_V^2}{D_V(q^2)} \cdot \frac{g_{\rho^0\pi^0\gamma}}{D_{\rho}(q'^2)} \nonumber \\
F^{\eta\pi^0\gamma}_{\omega}&=&ie\sum_{V=\omega, \phi \cdots} \frac{g_{V\omega\eta}}{\gamma_V}\cdot \frac{m_V^2}{D_V(q^2)} \cdot \frac{g_{\omega\pi^0\gamma}}{D_{\omega}(q'^2)} \nonumber \\
F^{\eta\pi^0\gamma}_{\phi}&=&ie\sum_{V=\omega, \phi \cdots} \frac{g_{V\phi\eta}}{\gamma_V}\cdot \frac{m_V^2}{D_V(q^2)} \cdot \frac{g_{\phi\pi^0\gamma}}{D_{\phi}(q'^2)} \nonumber \\
F^{\eta\eta\gamma}_{\rho}&=&ie\sum_{V=\rho, \rho', \rho'' \cdots} \frac{g_{V\rho^0\eta}}{\gamma_V}\cdot \frac{m_V^2}{D_V(q^2)} \cdot \frac{g_{\rho^0\eta\gamma}}{D_{\rho}(q'^2)} \nonumber \\
F^{\eta\eta\gamma}_{\omega}&=&ie\sum_{V=\omega,\phi, \cdots} \frac{g_{V\omega\eta}}{\gamma_V}\cdot \frac{m_V^2}{D_V(q^2)} \cdot \frac{g_{\omega\eta\gamma}}{D_{\omega}(q'^2)} \nonumber \\
F^{\eta\eta\gamma}_{\phi}&=&ie\sum_{V=\omega,\phi, \cdots} \frac{g_{V\phi\eta}}{\gamma_V}\cdot \frac{m_V^2}{D_V(q^2)} \cdot \frac{g_{\phi\eta\gamma}}{D_{\phi}(q'^2)} \ .
\eeqn

In the above expressions, the ellipsis in the sum over $V$ $s$-channel resonances include all possible radial excitations of vector mesons. For identical pseudoscalar mesons in the final state, one needs to exchange $q_1 \leftrightarrow q_2$ in the decay amplitudes, with the corresponding $q' \leftrightarrow q''$ two-particle momenta. Note that for non-identical particles ($\pi^0\eta$), the form factor for exchanged mesons are not given by the simple exchange of momenta because of the different isospin of $\pi^0$ (isovector) and $\eta$ (isoscalar) mesons.

\acknowledgements{}

 We are thankful to Michel Davier and Zhiqing Zhang for useful discussions and comments to this manuscript. We are grateful to Pablo Roig for comments. The work of GLC was supported by Ciencia de Frontera Conacyt (M\'exico) project No. 428218. JLGS is grateful to Conacyt for financial support through a Ph.D. scholarship.

\bibliography{bibfile}

\end{document}